\documentclass[a4paper,11pt]{article}

\usepackage{jheppub}

\usepackage{datetime}

\usepackage{amsmath,graphicx,amssymb,subfigure,bbm}
\usepackage[all]{xy}

\title{Multiple backreacted flavour branes}

\author[a]{Veselin G. Filev,}
\author[b]{Dimitrios Zoakos,}
\affiliation[a]{  School of Theoretical Physics, Dublin Institute for Advanced Studies, \\ 10 Burlington Road, Dublin 4, Ireland.}
\affiliation[b]{Centro de F\'isica do Porto \& 
Departamento de F\'isica e Astronomia, Faculdade de Ci{\^e}ncias \\ da Universidade do Porto, Rua do Campo Alegre 687, 4169--007 Porto, Portugal.}       
\emailAdd{vfilev@stp.dias.ie}\emailAdd{dimitrios.zoakos@fc.up.pt}

\abstract{We construct a novel supergravity background holographically dual to the
flavoured ${\cal N}=1$ Supersymmetric Yang-Mills theory. We consider flavours of different masses that produce spherical cavities with radii corresponding to the quark masses. 
The positive beta function blows at some large radial distance corresponding to the Landau pole of the theory.  
We explore the Wilson loop between two families of light and heavy quarks and observe a screening of the heavy light potential.}

\begin{document}
\maketitle
\flushbottom


\section{Introduction}

The gauge/gravity correspondence \cite{Maldacena:1997re} (for a resent pedagogical introduction see \cite{Ramallo:2013bua}) can potentially address many 
features of strongly coupled gauge theories. The original formulation of the duality relates ${\cal N}=4$ SYM with a string theory on AdS$_5\times S^5$.

An important extension of the duality, that initially was limited to adjoint degrees of freedom, is the inclusion of fundamental matter \cite{Karch:2002sh}. 
That is realised though flavour branes, extending along the holographic direction, occupying the gauge theory directions and wrapping some non-compact internal cycle 
(in order to promote a symmetry of the worldvolume to a global flavour symmetry). When the number of flavours is significantly less than the number of the colours, we are
in the limit that we can safely neglect their effect on the geometry and consider them as probes. On the contrary, when the number of flavours and colours become comparable,
a fully backreacted background needs to be constructed, since we cannot neglect anymore the backreaction of the flavour branes on the geometry.

In order to follow this path we need to solve the equations of motion that originate from a system that combines gravity plus brane sources. 
Since generically it is difficult to construct localised backreacted solutions we will follow a different approach  \cite{Bigazzi:2005md, Casero:2006pt} 
(see \cite{Nunez:2010sf} for a review and \cite{smearing} for some more smeared solutions), 
that delocalises  the branes' sources. The advantage of this approach is the replacement of the delta functions in the equations of motion
with continuous brane distribution functions. Instead of $U(N_f)$, which would be for the localised solution, 
the flavour symmetry of the dual gauge theory for the smeared solution is  $U(1)^{N_f}$. 
The smeared solutions are generally less supersymmetric but much simpler (and sometimes analytic as in the current paper) with respect to the localised.
On the field theory side the smeared construction is equivalent to the Veneziano limit, where both the number of colours and flavours are large but their ratio is fixed and finite. 

There are two types of backreacted flavours in a gravity setup. On one hand, when the flavours extend along the full range of the holographic coordinate, 
the solutions correspond to the addition of massless quarks. These backgrounds (for the D3/D7 case see  \cite{Benini:2006hh, Benini:2007gx}) generally possess a curvature 
singularity in the IR, which is the result of a high brane density close to the origin that all the flavours pass through. 
On the other hand, when the flavours do not reach the origin of the geometry, the solutions correspond to the addition of massive quarks.
Such a construction will remove the IR singularity (for the D3/D7 case with regular massive solutions  see \cite{Bigazzi:2008zt, Bigazzi:2008qq}).

In this paper we  move one step forward from the massive solution of \cite{Bigazzi:2008zt} in order to produce a new background that has multiple cavities of D7 flavour branes. 
This background will be supersymmetric, like the analog with one cavity, since kappa symmetry does not prevent us from adding a second (or multiple cavities).
We will test this new solution by calculating the beta function and the running of the coupling constant. Constraining to the case of two cavities, we perform 
the wilson loop computation that extends between them. This will reflect the forces between the quarks in a heavy light meson. 
This result extends previous computations in the probe limit (see \cite{Erdmenger:2007vj}) and confirms the screening effect of the flavours previously 
seen  in \cite{Bigazzi:2008zt, Bigazzi:2008qq} in a slightly different context.  


\section{Construction of the geometry}

The background consists of colour D3-branes and flavour D7-branes that extend along the radial direction and smear homogeneously over the transverse space.


\subsection{Ansatz and the BPS equations}

The ansatz for the metric (in the Einstein frame) is (inspired from \cite{Bigazzi:2009bk, Benini:2006hh})
\begin{equation} \label{10dmetric}
ds_{10}^2 = h^{-\frac{1}{2}}\Big[-dt^2 + dx_1^2+ dx_2^2+dx_3^2 \Big] + h^\frac{1}{2}
\Big[S^8F^2 d\sigma^2 + S^2 ds_{CP^2}^2 + F^2 (d\tau + A_{CP^2})^2 \Big] \, ,
\end{equation}
where the $CP^2$ metric has the following parametrisation 
\begin{eqnarray}
ds_{CP^2}^2&=&\frac{1}{4} d\chi^2+ \frac{1}{4} \cos^2 \frac{\chi}{2} (d\theta^2 +
\sin^2 \theta d\varphi^2) + \frac{1}{4} \cos^2 \frac{\chi}{2} \sin^2 \frac{\chi}{2}(d\psi + \cos \theta d\varphi)^2
\quad \& \nonumber \\ [3pt]
A_{CP^2}&=& \frac12\cos^2 \frac{\chi}{2}(d\psi + \cos \theta d\varphi)\,\,.
\label{cp2metric}
\end{eqnarray}
The range of the angles is $0\leq (\chi, \theta) \leq \pi$,  $0\leq \varphi, \tau < 2\pi$, $0\leq \psi< 4 \pi$.
The background is supplemented with a set of RR forms
\begin{eqnarray}  \label{RR}
&& F_5\,=\,\frac{Q_c}{h^2} \,dt \wedge dx_1 \wedge dx_2 \wedge dx_3 \wedge d\sigma\,+\,{\rm Hodge\,\,dual} \, ,
\nonumber \\
&& F_1\,= \, \displaystyle\sum_{\kappa =1}^{\eta}  Q_f^{\kappa} \, p_{\kappa}(\sigma) \, \theta(\sigma \, - \, \sigma_{\kappa}) \, (d\tau \, + \, A_{CP^2} ) \, ,
\end{eqnarray}
where the latter parametrises the presence of a smeared fundamental matter in the system. 
Explaining the symbols that appear in \eqref{RR}: $Q_f^{\kappa}$, $p_{\kappa}$ \& $\sigma_{\kappa}$ 
are the charge, the distribution function and the position of every cavity, respectively\footnote{The cavity charges $Q_f^{\kappa}$ sum to the full flavour charge
\begin{equation*}
\displaystyle\sum_{\kappa =1}^{\eta}  Q_f^{\kappa} \, = \, Q_f 
\end{equation*}
}.
As a result of the smearing, all the functions of the ansatz, $h,S,F,  \& \, \Phi$, depend only on the radial coordinate, while the function $p_{\kappa}(\sigma)$ depends on the brane embedding. We choose the usual frame for \eqref{10dmetric}
\begin{equation}
\arraycolsep=0.6cm
 \begin{array}{ll}
  e^0 \, = \,  h^{-{1 \over 4}} \, dt \, ,  & e^5 \, = \, \frac{1}{2} \, h^{1 \over 4} \, S \, d\chi \\ [3pt]
  e^1 \, = \,  h^{-{1 \over 4}} \, dx_1 \, ,  & e^6 \, = \, \frac{1}{2} \, h^{1 \over 4} \, S \, \cos\frac{\chi}{2} \, d\theta \\ [3pt]
  e^2 \, = \,  h^{-{1 \over 4}} \, dx_2 \, , & e^7 \, = \, \frac{1}{2} \, h^{1 \over 4} \, S \, \cos\frac{\chi}{2} \, \sin\theta \, d\phi \\ [3pt]
  e^3 \, = \,  h^{-{1 \over 4}} \, dx_3 \, ,  & e^8 \, = \,  \frac{1}{2} \, h^{1 \over 4} \, S \, \cos\frac{\chi}{2} \, \sin\frac{\chi}{2} 
 \left(d\psi \, + \,  \cos\theta \, d\phi\right)\\ [3pt]
  e^4 \, = \,  h^{1 \over 4} \, S^4 \, F \, d\sigma \, , & e^9 \, = \,  h^{1 \over 4} \, F \, \left(d\tau \, + \,  A_{CP^2}\right) \, . 
 \end{array}
\end{equation} 
The constants $Q_{c}$ and $Q_{f}$ are proportional to the number of colours and flavours
\begin{equation}
N_c = \frac{Q_c\, Vol(S_5)}{(2\pi)^4g_s \,\alpha'^2} \quad \& \quad
N_f = \frac{4\,Q_f\,Vol(S_5)}{2 \, \pi^2 \, g_s} \, .
\end{equation}
The killing spinor in the frame basis defined in \eqref{10dmetric} is \cite{Itsios:2013uya}
\begin{equation} \label{spinor}
\epsilon \, = \,  h^{-\frac{1}{8}}\, e^{-\frac{1}{2} \, i\, \sigma_2 \, \psi}
\,e^{-\frac{3}{2}\, i \, \sigma_2 \, \tau}\,\,\eta
\end{equation}
where $\eta$ is a constant spinor that satisfies the following set of projections
\begin{equation} \label{projections}
 i \, \Gamma_{0123}\,\sigma_2 \, \eta \, = \,  i \, \Gamma_{49}\,\sigma_2 \, \eta \, = \,  \eta \quad \& \quad 
i \, \Gamma_{58}\,\sigma_2 \, \eta \, = \, i \, \Gamma_{67}\,\sigma_2 \, \eta \, = \, - \, \eta \, .
\end{equation}
The analysis of the different components of the susy variations for the dilatino and the gravitino, leads to the following BPS system of first-order differential equations
\cite{Benini:2006hh, Bigazzi:2009bk, Itsios:2013uya}
\begin{eqnarray} 
& \partial_\sigma h \, = \, - \, Q_c  \, ,\quad \quad \quad 
\partial_\sigma F \, = \,  S^4 \, F 
\left[ 3 \, - \, 2 \, \frac{F^2}{S^2}\,  - \,  \frac{1}{2} \,  e^\Phi \, 
\displaystyle\sum_{\kappa =1}^{\eta}  Q_f^{\kappa} \, p_{\kappa}(\sigma) \, \theta(\sigma \, - \, \sigma_{\kappa})\right] \, ,&
\nonumber \\[5pt]
&\partial_\sigma S \, = \, S^3 \,  F^2  \, , \quad \quad \quad 
\partial_\sigma \Phi \, =  \, e^\Phi \, S^4 \, 
\displaystyle\sum_{\kappa =1}^{\eta}  Q_f^{\kappa} \, p_{\kappa}(\sigma) \, \theta(\sigma \, - \, \sigma_{\kappa}) \, .&
\label{massiveBPS}
\end{eqnarray}


\subsection{Solving the BPS system of equations} \label{sec:solvingBPS}

Combining the BPS equations \eqref{massiveBPS} with the equation for the profile of the embedding (see the kappa symmetry analysis in appendix \ref{app:kappa}) and 
the explicit expression for the brane distribution function (see appendix \ref{app:BDF}), we have a system of first order differential equations that 
satisfy the second order equations of motion. 
In order to solve this system we change variables from $\sigma$ to $\rho$ in the following way, $d \rho = S^4 d \sigma$. 
The embedding profile in the $\rho$-coordinate is obtained in the appendix  \ref{app:kappa} and it is 
\begin{equation} \label{embedding}
\chi_{\kappa} (\rho) \, = \, 2 \, \arcsin \frac{e^{\rho_{\kappa}}}{e^{\rho}} \, , 
\end{equation}
where $\rho_{\kappa}$ is the radius of the $\kappa$-th cavity. Combining \eqref{embedding} with \eqref{p_function} and \eqref{massiveBPS}, after changing 
variables from $\sigma$ to $\rho$, we end up with the following system of differential equations
\begin{eqnarray}
& \partial_\rho h \, = \, - \,Q_c \,  S^{-4} \, ,\quad \quad \quad 
\partial_\rho \log F \, = \, 
 3 \, - \, 2 \, \frac{F^2}{S^2}\,  - \,  \frac{1}{2} \,  e^\Phi \, 
\displaystyle\sum_{\kappa =1}^{\eta}  Q_f^{\kappa} \, \left(1\,- \,  \frac{e^{2 \rho_{\kappa}}}{e^{2 \rho}}\right)^2\, \theta(\rho \, - \, \rho_{\kappa}) \, ,&
\nonumber \\[5pt]
&S \, \partial_\rho S \, =  \,  F^2  \, , \quad \quad \quad 
\partial_\rho \Phi \, =  \, e^\Phi \, 
\displaystyle\sum_{\kappa =1}^{\eta}  Q_f^{\kappa} \, \left(1\,- \,  \frac{e^{2 \rho_{\kappa}}}{e^{2 \rho}}\right)^2\, \theta(\rho \, - \, \rho_{\kappa}) \, .&
\label{massiveBPS-2}
\end{eqnarray} 
Following the same analysis as in \cite{Itsios:2013uya}, we define a set of new fields $Z$, $U$ and $V$ as
\begin{equation} \label{definitions}
W(\rho) \, = \, S(\rho)^4 \, , \quad  
V(\rho) \, = \, \frac{F(\rho)^2}{S(\rho)^2} \, , \quad \& \quad 
Z(\rho)=e^{\Phi(\rho)} \, .
\end{equation}
The equations of motion for $W$, $V$ and $Z$ are
\begin{eqnarray}
\partial_{\rho}\log W(\rho)&=&4\, V\,  , \label{EOMr-W}\\  [3.5pt]
\partial_{\rho}\log V(\rho)&=&6\,(1-V)-p(\rho)\,Z\,  , \label{EOMr-V} \\  [3.5pt]
\partial_{\rho}\log Z(\rho)&=&p(\rho)\,Z \, ,  \label{EOMr-Z}
\end{eqnarray}
with 
\begin{equation}
p(\rho) \, \equiv \, \displaystyle\sum_{\kappa =1}^{\eta}  Q_f^{\kappa} \, \left(1\,- \,  \frac{e^{2 \rho_{\kappa}}}{e^{2 \rho}}\right)^2\, \theta(\rho \, - \, \rho_{\kappa}) \, .
\end{equation}
while the equation for $h$ is decoupled and will be solved in the end separately. 
The equation of motion for $Z$ can be easily integrated
\begin{equation}
Z^{-1} \, = \, 1 \, + \,  \displaystyle\sum_{\kappa =1}^{\eta}  Q_f^{\kappa} \,\left(f_{\kappa}(\rho_*) \, - f_{\kappa}(\rho_{\kappa})\right) \, - \, 
 \displaystyle\sum_{\kappa =1}^{\eta}  Q_f^{\kappa} \,\left(f_{\kappa}(\rho) \, - f_{\kappa}(\rho_{\kappa})\right) \, \theta(\rho \, - \, \rho_{\kappa}) \, ,  
\label{Solution-Z}
\end{equation}
where $\rho_*$ is a radial UV scale and we fix the constant of integration in a way that $Z(\rho_*) =1$. 
The function $f_{\kappa}(\rho)$ is defined as follows
\begin{equation} \label{def-f}
f_{\kappa}(\rho) \, = \, \rho \, + \,  \frac{e^{2 \rho_{\kappa}}}{e^{2 \rho}} \, - \, \frac{1}{4} \,  \frac{e^{4 \rho_{\kappa}}}{e^{4 \rho}}\, .
\end{equation}
The function $Z$, that is related to the dilaton, has a pole which is determined by the following relation
\begin{equation}
1 \, + \,  \displaystyle\sum_{\kappa =1}^{\eta}  Q_f^{\kappa} \,\left(f_{\kappa}(\rho_*) \, - f_{\kappa}(\rho_{LP})\right) \, = \, 0  \, . \label{LP}
\end{equation}
The dilaton diverges at $\rho_{LP}$, which is the energy scale corresponding to the Landau pole of the theory. 
This is the outcome of the positive contribution of the flavour to the beta function of the theory. 

\noindent
Proceeding as in \cite{Itsios:2013uya}, we combine (\ref{EOMr-W}), (\ref{EOMr-V}) and (\ref{EOMr-Z}) in obtaining an equation for $V$
\begin{equation} \label{EOM-VtoWZ}
\partial_{\rho}\log(Z\,V\,W^{3/2}) \, = \, 6  \quad \Rightarrow \quad V(\rho) \, = \, \frac{c_V \, e^{6\rho}}{Z(\rho) \, W(\rho)^{3/2}} \, . 
\end{equation}
Substituting \eqref{EOM-VtoWZ} into \eqref{EOMr-W} and using \eqref{Solution-Z}, 
we obtain a differential equation for $W$, which we solve as follows
\begin{equation}
W(\rho) \, = \, \alpha'^2\, e^{4\rho}\,\Bigg[\displaystyle\sum_{\kappa =1}^{\eta}  Q_f^{\kappa} \,\left(f_{\kappa}(\rho_{LP}) \, - f_{\kappa}(\rho_{\kappa})\right) \, - \, 
 \displaystyle\sum_{\kappa =1}^{\eta}  Q_f^{\kappa} \,\left(A_{\kappa}(\rho) \, - A_{\kappa}(\rho_{\kappa})\right) \, \theta(\rho \, - \, \rho_{\kappa})\Bigg]^{2/3} \ ,\label{Solution-W}
\end{equation}
where we have fixed the constant of integration in complete analogy to \cite{Benini:2006hh,Itsios:2013uya} and used \eqref{LP} . 
The function $A_{\kappa}(\rho)$ is defined as follows
\begin{equation} \label{def-A}
A_{\kappa}(\rho) \, = \, \rho \, + \,  \frac{3}{2}\,\frac{e^{2 \rho_{\kappa}}}{e^{2 \rho}} \, - \, \frac{3}{4} \,  \frac{e^{4 \rho_{\kappa}}}{e^{4 \rho}} 
\, + \, \frac{1}{6} \,  \frac{e^{6 \rho_{\kappa}}}{e^{6 \rho}}\, .
\end{equation}
The corresponding solution for $V(\rho)$ is
\begin{equation}
V(\rho) \, = \, \frac{\displaystyle\sum_{\kappa =1}^{\eta}  Q_f^{\kappa} \,\left(f_{\kappa}(\rho_{LP}) \, - f_{\kappa}(\rho_{\kappa})\right) \, - \, 
 \displaystyle\sum_{\kappa =1}^{\eta}  Q_f^{\kappa} \,\left(f_{\kappa}(\rho) \, - f_{\kappa}(\rho_{\kappa})\right) \, \theta(\rho \, - \, \rho_{\kappa})}
{ \displaystyle\sum_{\kappa =1}^{\eta}  Q_f^{\kappa} \,\left(f_{\kappa}(\rho_{LP}) \, - f_{\kappa}(\rho_{\kappa})\right) \, - \, 
 \displaystyle\sum_{\kappa =1}^{\eta}  Q_f^{\kappa} \,\left(A_{\kappa}(\rho) \, - A_{\kappa}(\rho_{\kappa})\right) \, \theta(\rho \, - \, \rho_{\kappa})}\, . \label{Solution-V}
\end{equation}
Combining \eqref{Solution-Z}, \eqref{Solution-W} \& \eqref{Solution-V} together
with \eqref{definitions} it is possible to obtain expressions for the $\Phi, F$ \& $S$ as functions of $\rho$.
These expressions are
\begin{eqnarray} \label{Solution-h-phi-S-F}
&& e^{- \Phi(\rho)} \, = \, \displaystyle\sum_{\kappa =1}^{\eta}  Q_f^{\kappa} \,\left(f_{\kappa}(\rho_{LP}) \, - f_{\kappa}(\rho_{\kappa})\right) \, - \, 
 \displaystyle\sum_{\kappa =1}^{\eta}  Q_f^{\kappa} \,\left(f_{\kappa}(\rho) \, - f_{\kappa}(\rho_{\kappa})\right) \, \theta(\rho \, - \, \rho_{\kappa}) \, , 
\nonumber \\ [5pt]
&&  S(\rho)  \, =  \,  \sqrt{\alpha'}  \, e^{\rho}\,\Bigg[ \displaystyle\sum_{\kappa =1}^{\eta}  Q_f^{\kappa} \,\left(f_{\kappa}(\rho_{LP}) \, - f_{\kappa}(\rho_{\kappa})\right) \, - \, 
 \displaystyle\sum_{\kappa =1}^{\eta}  Q_f^{\kappa} \,\left(A_{\kappa}(\rho) \, - A_{\kappa}(\rho_{\kappa})\right) \, \theta(\rho \, - \, \rho_{\kappa}) \Bigg]^{1/6} \, , 
\nonumber\\ [5pt] 
&&  F(\rho)  \, =  \,  \sqrt{\alpha'} \, \, e^{\rho}\, 
\Big[\displaystyle\sum_{\kappa =1}^{\eta}  Q_f^{\kappa} \,\left(f_{\kappa}(\rho_{LP}) \, - f_{\kappa}(\rho_{\kappa})\right) \, - \, 
 \displaystyle\sum_{\kappa =1}^{\eta}  Q_f^{\kappa} \,\left(f_{\kappa}(\rho) \, - f_{\kappa}(\rho_{\kappa})\right) \, \theta(\rho \, - \, \rho_{\kappa}) \Big]^{1/2}\, \times 
\nonumber \\ [5pt]
 && \quad \quad \quad \times  \Bigg[  \displaystyle\sum_{\kappa =1}^{\eta}  Q_f^{\kappa} \,\left(f_{\kappa}(\rho_{LP}) \, - f_{\kappa}(\rho_{\kappa})\right) \, - \, 
 \displaystyle\sum_{\kappa =1}^{\eta}  Q_f^{\kappa} \,\left(A_{\kappa}(\rho) \, - A_{\kappa}(\rho_{\kappa})\right) \, \theta(\rho \, - \, \rho_{\kappa})\Bigg]^{-1/3} \, .
\end{eqnarray}


\section{Properties of the solution}

In this section we will probe the new solution by calculating several physical observables. 
In order to simplify the analysis most of the time we will restrict to cases with two cavities.


\subsection{Calculation of the quark masses}

The constituent mass $m_q$ of the dynamical quarks is related to the radial distance $\rho_q$ of each one of the multiple cavities of the solution. 
As usual it is proportional to the energy of a straight string stretched along the radial direction, from the origin of the geometry at $\rho \to - \infty$
to the position of every spherical cavity. For the physical mass of the massive flavours corresponding to the $\kappa$-th family, using the Nambu-Goto  action we obtain:
\begin{eqnarray}
m_{q\,\kappa}  & = &  \frac{1}{2\,\pi\,\alpha'}\,\int\limits_{-\infty}^{\rho_{\kappa}}\,d\rho\,\sqrt{-G_{00}^{s}\,G_{\rho\rho}^{s}}\,=\,\frac{1}{2\,\pi\,\alpha'}\,\int\limits_{-\infty}^{\rho_{\kappa}}\,d\rho\,e^{\frac{1}{2}\Phi(\rho)}\,F(\rho)\ .
\end{eqnarray}
Where the functions $\Phi(\rho)$ and $F(\rho)$ are defined in equation (\ref{Solution-h-phi-S-F}) and $\rho_{\kappa}$ is the radius to the cavity. When there is just one cavity at $\rho = \rho_q$ we re-obtain the result of \cite{Bigazzi:2008zt}:
\begin{eqnarray}
\nonumber \\ [4pt]
m_{q} & = &  \frac{1}{2 \pi \sqrt{\alpha'}} \, \frac{e^{\rho_q}}{Q_f^{1/3}} \, \Big[ f_{\kappa}(\rho_{LP}) \, -  f_{\kappa}(\rho_q) \Big]^{-1/3} \, , 
\end{eqnarray}
where $Q_f$ \& $\rho_q$ are the charge and the radial position of the spherical cavity respectively. As it was observed in  \cite{Bigazzi:2008zt}, 
in the limit $\rho_q \to \rho_{LP}$ the mass diverges, but this is consistent with the fact that we do not trust the supergravity solution in this limit since the dilaton blows up.

Going one step further we will focus on the case with two spherical cavities and calculate the masses of the two quarks, one ``light" and one ``heavy". 
The mass of the ``light"  quark, that corresponds to a string stretching from the origin until the first cavity $\rho_1$ (we have assumed that 
$\rho_2>\rho_1$) is given by the following 
expression
\begin{equation} \label{lightQuark}
m_{light} \, = \, \frac{1}{2 \pi \sqrt{\alpha'}} \, {e^{\rho_1}} \, 
\left[\displaystyle\sum_{\kappa =1}^{2}  Q_f^{\kappa} \,\left(f_{\kappa}(\rho_{LP}) \, - f_{\kappa}(\rho_{\kappa})\right) \right]^{-\frac{1}{3}} \, .
\end{equation}
where $Q_f^1$ \& $Q_f^2$ are the flavour charges of every cavity.  The calculation for the mass of the second quark can split in two pieces.
The first one is going to give the mass of the ``light"  quark in \eqref{lightQuark}, while the second is going to essentially give the difference between the masses.
The explicit formula is the following
\begin{equation} \label{massdif-v1}
m_{heavy} \, = \, m_{light} \, +\, \frac{1}{2 \pi \sqrt{\alpha'}} \, \int\limits_{\rho_1}^{\rho_{2}} \, e^{\rho} \, 
\left[\displaystyle\sum_{\kappa =1}^{2}  Q_f^{\kappa} \,\left(f_{\kappa}(\rho_{LP}) \, - f_{\kappa}(\rho_{\kappa})\right) \, - \, 
 Q_f^{1} \,\left(A_{\kappa}(\rho) \, - A_{\kappa}(\rho_{1})\right) \right]^{-\frac{1}{3}} .
\end{equation}
Since the above integral cannot be performed analytically, we will expand for $Q_f^1\ll Q_f^2$ and perform the integral order by order in the expansion.
In order to present the result in a compact form we introduce the following auxiliary function 
\begin{equation} \label{def-g}
g(\rho) \, = \, \frac{e^{\rho}}{e^{\rho_1}} \, \left( \frac{7}{6} \, - \, \rho \, + \rho_{LP} \right) \, + \, \frac{1}{30} \, \frac{e^{5 \rho_1}}{e^{5 \rho}} \, 
+ \, \frac{3}{2} \, \frac{e^{\rho_1}}{e^{\rho}} \, -  \, \frac{1}{4} \, \frac{e^{3 \rho_1}}{e^{3 \rho}} \, + \, \frac{e^{\rho \,+\, \rho_1}}{e^{2 \rho_{LP}}}
\, - \, \frac{1}{4}\frac{e^{\rho \,+\, 3 \rho_1}}{e^{4 \rho_{LP}}}  \,. 
\end{equation}
Using \eqref{def-g} it is easier to derive the following formula, which gives the ratio of the mass difference and the mass of the ``light" quark as a series
expansion in $\epsilon \equiv \frac{Q_f^1}{Q_f^2}$.
\begin{eqnarray}
 \frac{m_{heavy} \, - \, m_{light}}{m_{light}} \,& =& \, \left( \frac{e^{\rho_2}}{e^{\rho_1}}  \, - \, 1 \right) \,  \times
\\ 
&&  \Bigg[ 1 \, + \, \frac{\epsilon}{3} \, \frac{f_{\kappa}(\rho_{LP}) - f_{\kappa}(\rho_1)}{f_{\kappa}(\rho_{LP}) - f_{\kappa}(\rho_2)}
\left( 1 \, - \, \frac{e^{\rho_1}}{e^{\rho_2} \, - \, e^{\rho_1}} \, \frac{g(\rho_2) \, - \,  g(\rho_1)}{f_{\kappa}(\rho_{LP}) - f_{\kappa}(\rho_1)}\right) \Bigg] \, .
\nonumber
\end{eqnarray}
For  $\epsilon =0$ the above formula gives zero for the mass difference, since in this case $\rho_2 = \rho_1$.


\subsection{Running of the coupling constant and the beta function}

\begin{figure}[h] 
   \centering
   \includegraphics[width=7cm]{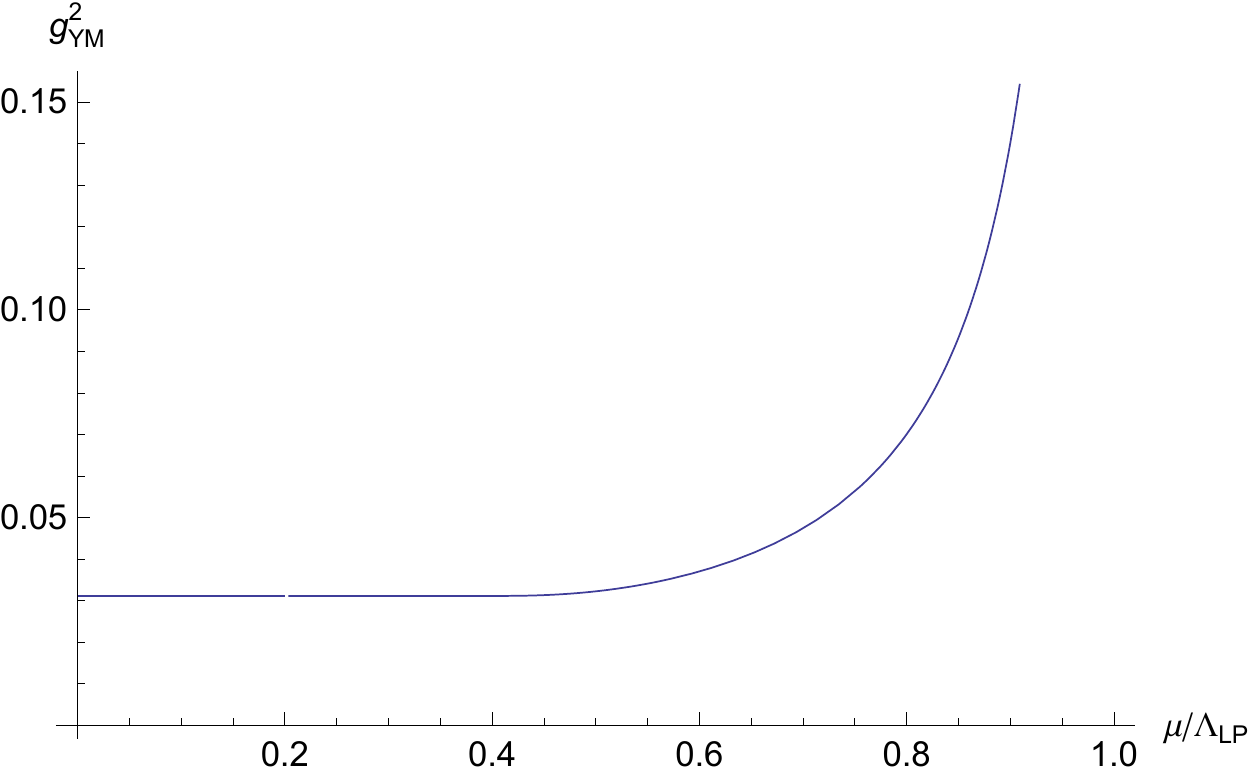}
    \includegraphics[width=7cm]{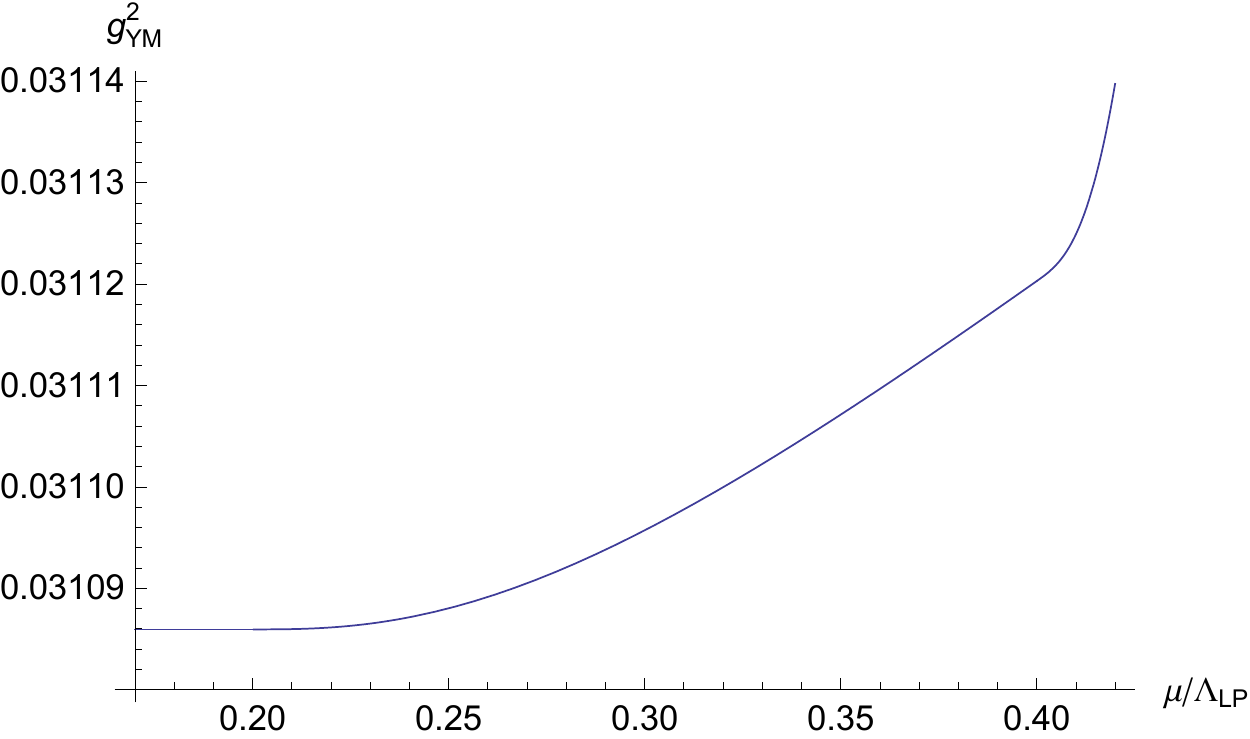}
    \includegraphics[width=7cm]{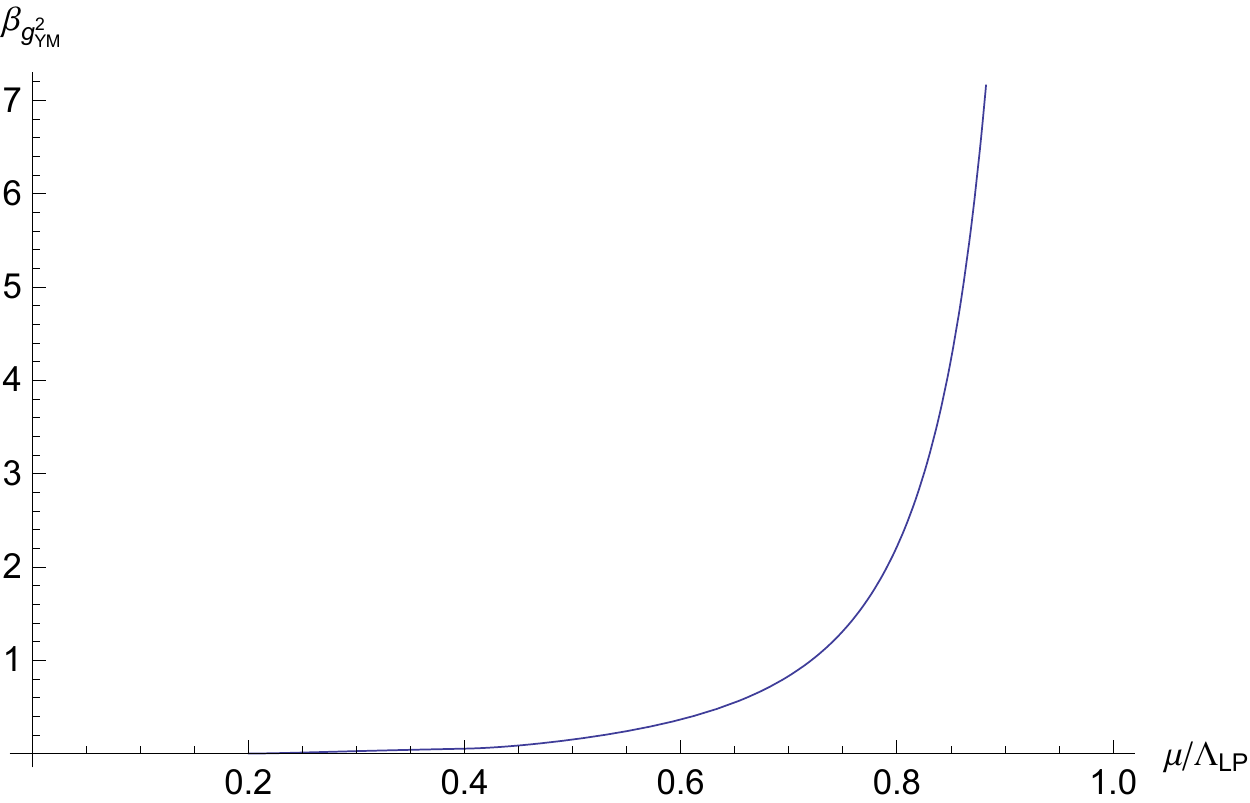}
     \includegraphics[width=7cm]{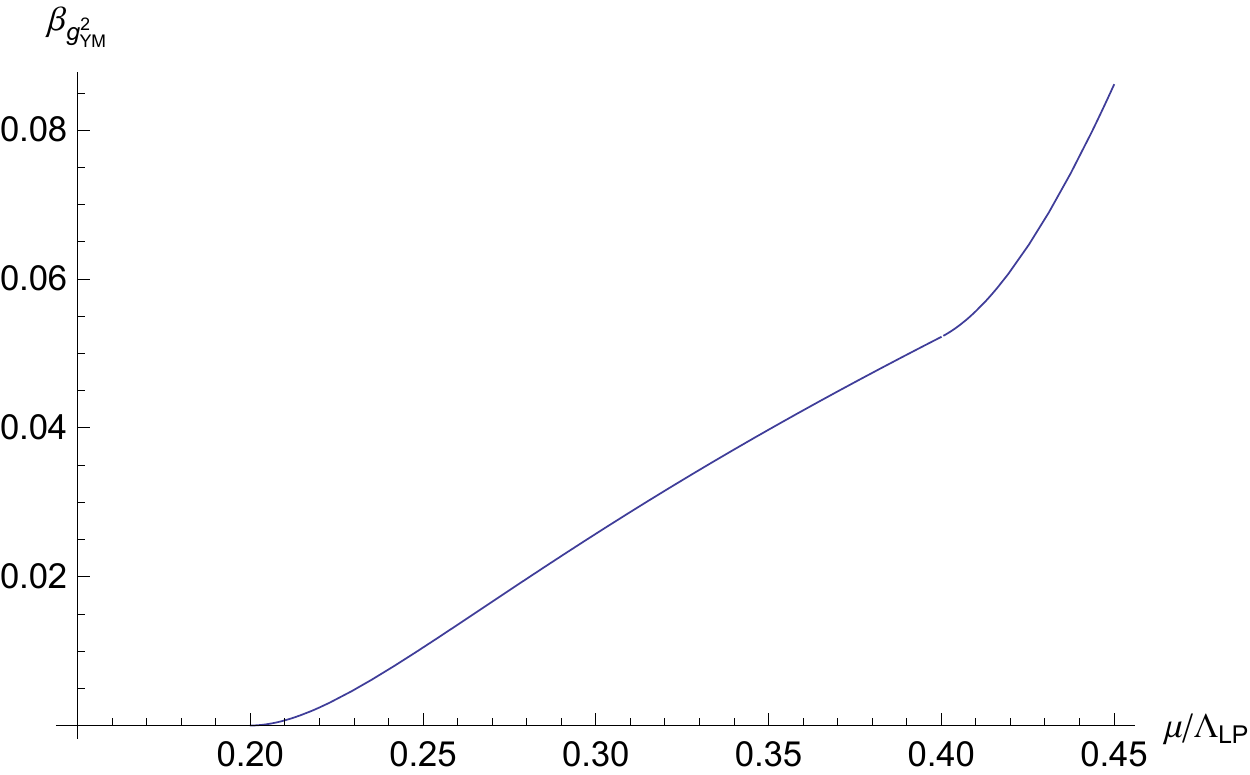}
   \caption{Plots for the dependence on the energy scale for both the coupling constant ${g_{YM}^2}$ and the beta function $\beta_{g_{YM}^2}$. 
    The two plots on the top \&  bottom right are magnifications of the corresponding plots on the left, in order to illustrate the effect of the presence of the cavities.}
   \label{fig:1}
\end{figure}

In this section we will study the Yang-Mills coupling $g_{YM}^2$ of the dual field theory and illustrate the dependence on the number of flavours distributed in more
than one cavity (for concreteness we will focus on the case of two cavities). The actual expression relating the coupling with the dilaton is
\begin{equation}\label{YM-coupling}
g_{YM}^2\,= \, 4\,\pi\,g_s\,e^{\Phi(\rho)} \, .
\end{equation}
In order to study the  running of $g_{YM}^2$ with the energy scale $\mu$ we need to specify the precise  radius/energy relation. 
We will use the  radius/energy relation that corresponds to the smearing of massless flavour D7-branes \cite{Benini:2006hh, Bigazzi:2008zt, Itsios:2013uya}
\begin{equation}\label{radius/energy}
\rho \, = \, \rho_{LP} \, + \, \log\frac{\mu}{\Lambda_{LP}} \quad \Rightarrow  \quad
\rho_{q_i} \, = \, \rho_{LP} \, + \, \log\frac{m_{q_i}}{\Lambda_{LP}}  \, , 
\end{equation}
an expression which is expected to be valid also for the smearing of massive flavour D7-branes 
(at least as a good approximation), when the flavours are close to the IR (corresponding to very light quarks).
Using the radius/energy relation (\ref{radius/energy}) we can study the running of $g_{YM}^2$ with the energy scale $\mu$.
Focusing again to the case of two cavities corresponding to ``light" and ``heavy" quarks we have
\begin{eqnarray} \label{coupling}
&& \frac{4 \pi g_s}{g_{YM}^2}\, = \, - \, \displaystyle\sum_{\kappa =1}^{2}  Q_f^{\kappa} \, \Big[ \frac{3}{4} \, + \, \log \frac{m_{\kappa}}{\Lambda_{LP}} \, - \, 
\left(\frac{m_{\kappa}}{\Lambda_{LP}}\right)^2 \, + \, \frac{1}{4}\, \left(\frac{m_{\kappa}}{\Lambda_{LP}}\right)^4 \Big]
\nonumber \\ [5pt]
&& - \, 
 \displaystyle\sum_{\kappa =1}^{2}  Q_f^{\kappa} \, \left[ - \,  \frac{3}{4} \, - \, \log \frac{m_{\kappa}}{\mu} \, +\, 
\left(\frac{m_{\kappa}}{\mu}\right)^2 \, - \, \frac{1}{4}\, \left(\frac{m_{\kappa}}{\mu}\right)^4 \right] \,\theta(\mu \, - \, m_{\kappa})  \, .
\end{eqnarray}
Moving one step forward (and  using again \eqref{radius/energy}) we can calculate the corresponding beta function $\beta_{g_{YM}^2}$, 
in order to determine its dependence on the presence of the different cavities.
Substituting the multicavity solution \eqref{Solution-h-phi-S-F} on the actual expression for the beta function
\begin{equation}
\beta_{g_{YM}^2}\,=\,\frac{\partial\,g_{YM}^2}{\partial\,(\log\frac{\mu}{\Lambda_{LP}})} =4\,\pi\,g_s\,e^{\Phi(\rho)}\,\Phi'(\rho)\, ,
\end{equation}
we get a very lengthy and non-illuminating expression that we will not write explicitly. Instead in figure \ref{fig:1} we provide the plots for the running 
of both the coupling constant and the beta function with respect to the energy scale. 

Plotting $g_{YM}^2$  and $\beta_{g_{YM}^2}$ for different values of both the quark masses and the charges of the spherical cavities, one can see that the 
corresponding quantity grows at all energy scales.
At the  energy scale (radial distance) that a cavity is located there is change on the slope of the corresponding quantity, and the change is related to the number 
 of flavours of the cavity. The more flavours we put, the more severe the change of the slope is.


\subsection{Wilson loops between cavities}

In the previous sections we constructed the gravity dual of ${\cal N} =4$ SYM with a large number of backreacted dynamical flavours, distributed in a series of multiple cavities. 
In order to study the multicavity effect on the perturbative dynamics we will probe this background with a quark-antiquark pair. 
We will constrain to the case of two cavities and study the embedding of a static string with one edge on the outer and the other edge on the inner cavity.\footnote{To this end we introduce two additional probe D7-branes wrapping contracting cycles in the internal directions of the geometry, and reaching minimal distances from the origin at each cavity. The ends of the strings are attached to each brane.} 
In this way we will study the forces between the quarks of a heavy-light meson.  So far in the literature there are studies of the flavour effects of a single cavity on the Wilson loop 
(see \cite{Bigazzi:2008zt} \& \cite{Bigazzi:2008qq}) or studies of Wilson loops for heavy-light mesons but in geometries without backreacting flavour (see \cite{Erdmenger:2007vj}).
In the current paper we will cover the full parametric space by studying heavy-light Wilson loops in a fully backreacted background.  

\begin{figure}[t] 
   \centering
   \includegraphics[width=13cm]{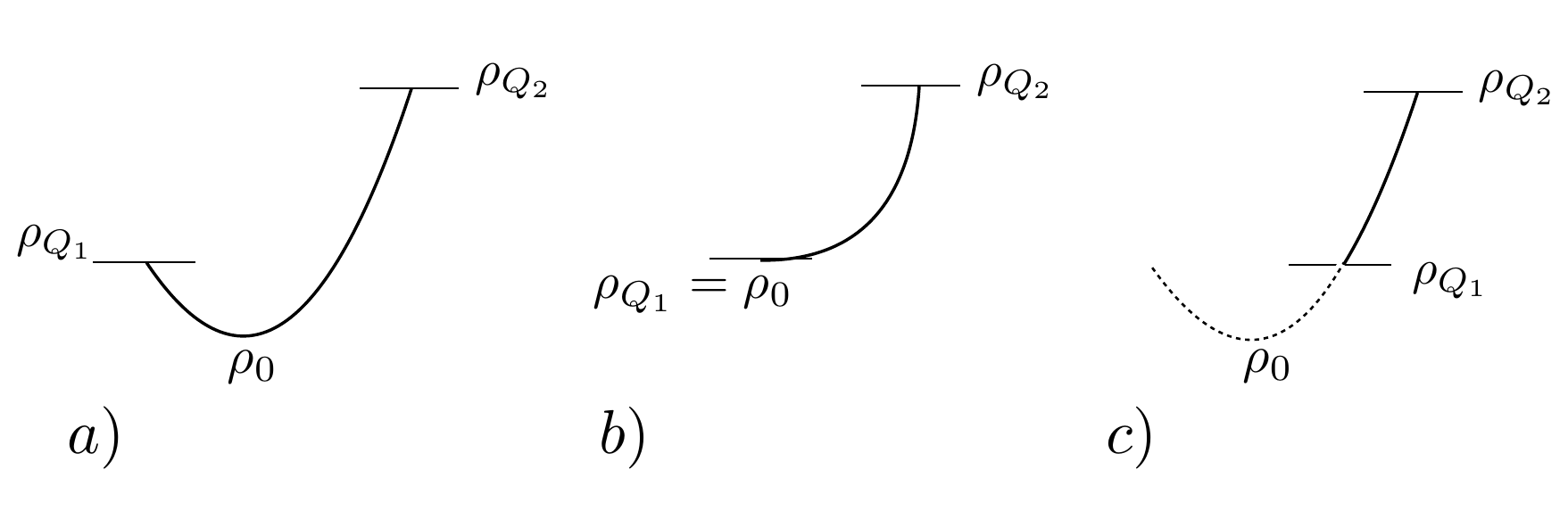}
   \caption{ Sketches of the different classes of string embeddings. Plot a) represents string embedding, which reaches a minimal radial distance between the end points, it is realised for sufficiently large separations of the quarks (the sting's endpoints). Plot b) represents a critical embedding when the position of the minimum is reached at the lower end of the string. Plot c) represents monotonic string embeddings, although the minimum is not realised the parameter $\rho_0$ still characterises the embedding.}
   \label{fig:2}
\end{figure}

The string in the bulk of the geometry bends and reaches a minimal radial distance $\rho_0$. 
The Minkowski separation $L$ between the quarks and the total energy of the system both depend on $\rho_{0}$ and manipulating those relations 
we can deduce the $\bar Q Q$ potential as a function  of the separation $L$.   As usual the open string embedding is 
\begin{equation}
t \, = \, \tau, \quad y \, = \, \sigma, \quad \rho \, = \, \rho(y) \, ,
\end{equation}
where $y \in [-L/2,L/2]$ is a Minkowski direction. The string action is 
\begin{equation}
S \, = \, - \, \frac{1}{2\pi\alpha'} \, \int dt dy  \sqrt{ H(\rho)^2\, + \, G(\rho)^2 \, (\partial_y \rho)^2 }\, ,  
\end{equation}
where the functions $F$ and $G$ are defined as follows 
\begin{equation}
H \, \equiv \,  e^{{\Phi\over2}} \, h^{-1/2}\, , \quad \quad  
G \, \equiv \, e^{{\Phi\over2}} \, F \,.
\end{equation}
Following the standard references (see \cite{Maldacena:1998im} \& \cite{Rey:1998ik}) we can write the string length and energy\footnote{When we use the subindex 
$0$ we refer to a quantity evaluated at  $\rho=\rho_0$}
\begin{eqnarray}\label{LandE}
 L(\rho_0)&=&
\int_{\rho_0}^{\rho_{Q_1}} \frac{G \, H_{0}}{H \, \sqrt{H^2\, - \, H_{0}^2}}\, d\rho \, + \,  
\int_{\rho_0}^{\rho_{Q_2}} \frac{G \, H_{0}}{H \, \sqrt{H^2 \, - \, H_{0}^2}} \, d\rho\,,  
\nonumber \\ 
E (\rho_0)&=& \frac{1}{2\pi\alpha'} \Biggl[
\int_{\rho_0}^{\rho_{Q_1}} \frac{G \, H}{\sqrt{H^2 \, - \, H_{0}^2}} \, d\rho \, + \, 
\int_{\rho_0}^{\rho_{Q_2}} \frac{G \, H}{\sqrt{H^2 \, - \, H_{0}^2}} \, d\rho  \Bigg] \, ,  
\end{eqnarray}
where $\rho_{Q_1}$ \& $\rho_{Q_2}$ are the radial positions of the light and heavy quark branes, respectively. In fact equations (\ref{LandE}) describe only one class of open string embeddings, when the embedding reaches a minimal radial distance between the two edges of the string (figure \ref{fig:2}$a$). These embeddings are realised when the quarks (attached to the probe flavour branes) are sufficiently separated. Decreasing the separation of the quarks leads to a configuration for which the minimum of the string coincides with its lower end point (figure \ref{fig:2}$b$). Configurations with lower separations are possible only for monotonic string embeddings (figure~\ref{fig:2}$c$). In this case the expressions for the string length and energy (assuming $\rho_{Q_1}<\rho_{Q_2}$) are modified to:
\begin{eqnarray}\label{LandE-zero}
 L(\rho_0)&=&
\int_{\rho_{Q_1}}^{\rho_{Q_2}} \frac{G \, H_{0}}{H \, \sqrt{H^2\, - \, H_{0}^2}}\, d\rho \,,  
\nonumber \\ 
E (\rho_0)&=& \frac{1}{2\pi\alpha'} \,
\int_{\rho_{Q_1}}^{\rho_{Q_2}} \frac{G \, H}{\sqrt{H^2 \, - \, H_{0}^2}} \, d\rho \,.
\end{eqnarray}
Note that the quantities with subindex $0$ in equation (\ref{LandE-zero}) are still evaluated at a radial distance $\rho_0$, which although not reached by the string, enters as a parameter characterising the shape of the embedding. Generally $\rho_0 \leq\rho_{Q_1}<\rho_{Q_2}$, and at $\rho_0 =\rho_{Q_1}$ one has the critical embedding from figure \ref{fig:2}$b$. In the limit $\rho_0\to-\infty$ a configuration with radially stretched string, corresponding to a bound state of heavy and light quarks, is realised.  
Another quality of interest is the potential $V$ between the quarks, which can be obtained by subtracting from the energy $E$ of the configuration the total mass of the quarks:
\begin{eqnarray}
V \, = \, E \, - \, (m_{light} \, + \, m_{heavy})\ ,
\end{eqnarray}
where $m_{light}$ and $m_{heavy}$ are given by (\ref{lightQuark}) and (\ref{massdif-v1}), and the energy $E$ is given either by (\ref{LandE}) or (\ref{LandE-zero}).

To study the effect of the dynamical flavours on the quark potential we have to fix a comparison scheme, namely we have to decide which quantities we will keep fixed while changing the number of flavours. The first choice that we will consider is to keep the radii of the cavities $\rho_i$ and the UV scale $\rho_*$ (defined below equation (\ref{Solution-Z})) fixed. The motivation for this scheme is the relation of the radii of the cavities to the bare quark masses. Indeed, in the limit of vanishing number of backreacted flavours, the fiducial embeddings determining the density of the smeared flavour branes describe probe branes in pure AdS$_5\times S^5$ space-time, and the parameters $e^{\rho_i}$ are exactly the bare masses of these branes. Therefore, to zeroth order in the number of backreacted flavours the radii of the cavities correspond to the bare mass parameters of the probe branes. The exact relation between the bare masses and the radii becomes more involved once back reaction is taken into account and we refer the reader to refs. \cite{Filev:2011mt,Erdmenger:2011bw,Magana:2012kh} for details. For the purpose of our study it is sufficient to know that keeping the radii of the cavities fixed corresponds to keeping the bare masses of the quarks fixed. The choice to keep the UV cutoff $e^{\rho_*}$ fixed is related to the fact that the position of the Landau pole should be allowed to change as the number of dynamical flavours is changed to ensure that the limit of zero number of backreacted flavours can be taken properly. 

\begin{figure}[h] 
   \centering
   \includegraphics[width=10cm]{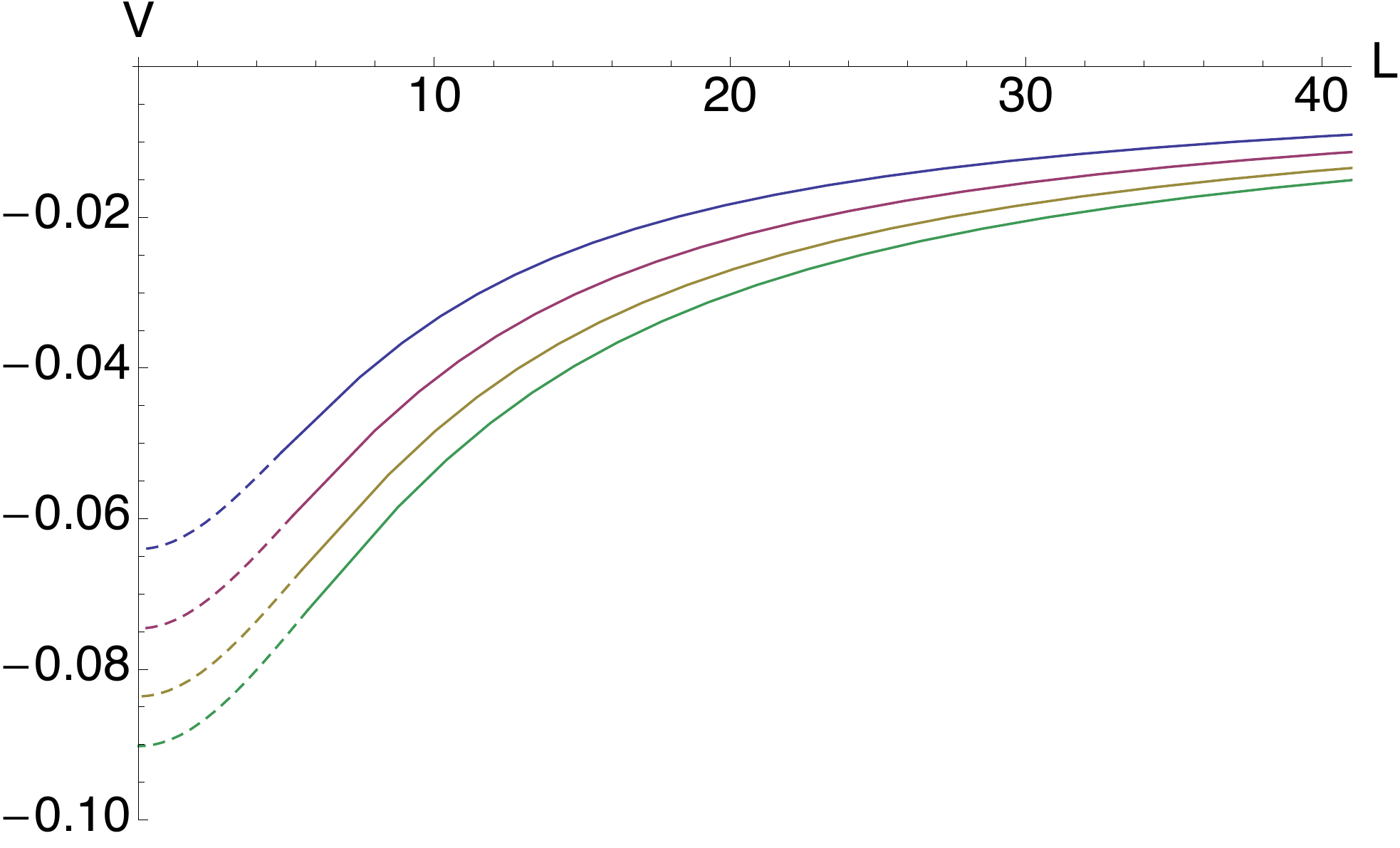}
   \caption{Plots of the heavy--light quark potential (in units of $(2\pi\sqrt{\alpha'})^{-1}$) for mass parameters (position of the cavities) $\rho_{q1} =-3,\,\rho_{q2} =-1$, and for equal numbers of heavy and light quarks ($Q_f^1=Q_f^2$). The total density of the flavour branes parameter $Q_f = Q_f^1+Q_f^2$ from the top to the bottom is: $1,\,1/2,\,1/4,\,1/8$. One can see that as the quark density is increased the quark potential becomes more shallow due to the screening of the dynamical flavours. }
   \label{fig:3}
\end{figure}

A plot of the quark potential $V$ versus the distance between the quarks $L$ for various number of dynamical flavours is presented in figure \ref{fig:3}. The plots are for mass parameters (related to the radial position of the cavities) $\rho_{q1} =-3,\,\rho_{q2} =0$, for UV cut-off parameter $\rho_*=2$, and for equal numbers of heavy and light quarks ($Q_f^1=Q_f^2$). The total density of the flavour branes parameter $Q_f = Q_f^1+Q_f^2$ from the top to the bottom is: $1,\,1/2,\,1/4,\,1/8$ and one can see that as the quark density is increased the quark potential becomes more shallow due to screening of the dynamical flavours. The solid segments of the curves correspond to configurations with realised minima in figure \ref{fig:3}$a$.
At large distance the potential follows the Coulomb law something that it is expected since the theory without flavours is conformal. On the contrary the dashed segments correspond to short string configurations (figure \ref{fig:3}$c$). One can see that as the separation $L$ approaches zero the potential $V$ is approximately harmonic. The transition from Coulomb to harmonic potential takes place near the critical configuration (figure \ref{fig:3}b), where the potential has an inflection point, and it is a smooth crossover. The value of the potential at $L=0$ corresponds to the binding energy of the heavy-light bound state. As expected one can see that it decreases as the number of dynamical flavours increases. 

\begin{figure}[h] 
   \centering
   \includegraphics[width=10cm]{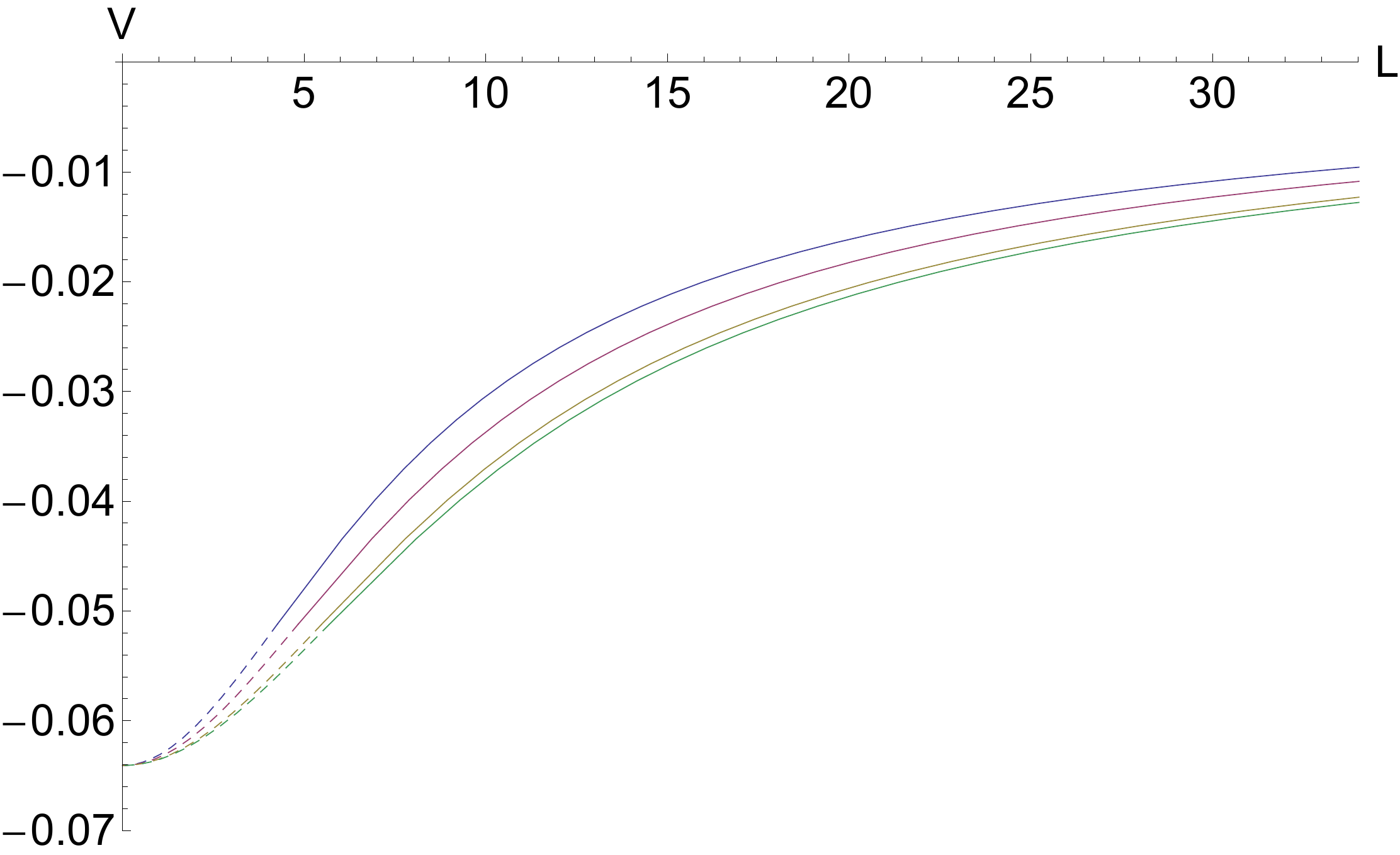}
   \caption{Plots of the heavy-light potential at fixed physical masses and fixed effective Yang-Mills coupling (at the outer cavity). From top to bottom the curves correspond to flavour brane density parameters $Q_f^1=Q_f^2=1,\,1/2\,,1/8\,,1/32$. One can see the screening effect of the dynamical flavours while at $L=0$ all curves start from the same value, since the binding energy remains fixed.}
   \label{fig:4}
\end{figure}

As we discussed above there are other comparison schemes that one could use. It is natural to compare the theories with different number of dynamical flavours when the infrared parameters of the theory, such as the physical masses and the low energy effective Yang-Mills coupling (given by equation (\ref{YM-coupling})) are kept fixed. Keeping the physical masses determines two of the parameters of the geometry. The third one could be the Yang-Mills coupling  either at the light quark energy scale (at the inner cavity) or at the heavy quark energy scale (the outer cavity). Intriguingly  if we fix the Yang-Mills coupling at the inner cavity we do not observe any screening as the number of flavours is changed. For different flavours the potential remains the same as the upper curve in figure \ref{fig:3}. On the contrary if we keep the physical masses and the effective Yang-Mills coupling at the outer cavity fixed we do observe a screening effect. The resulting plots are presented in figure \ref{fig:4}. From top to bottom the curves correspond to flavour brane density parameters $Q_f^1=Q_f^2=1,\,1/2\,,1/8\,,1/32$. Note also that at $L=0$ the curves have the same binding energy $V(0)$. The reason is that $L=0$ corresponds to the limit $\rho_0\to\infty$, when $H_0$ in equation (\ref{LandE-zero}) vanishes and the expression for the energy reduces to the expression for the difference of the masses $E(0)=m_{heavy}-m_{light}$ (see equation (\ref{massdif-v1})). For the binding energy $V(0)$ we obtain $V(0)=E(0)-(m_{heavy}+m_{light}) = -2\,m_{light}$ and hence it remains fixed if the physical mass of the light quark is kept fixed.

%
%

%

%


\section{Conclusions and outlook}

In this paper we constructed a novel supergravity background holographically dual to a flavoured ${\cal N}=1$ Supersymmetric Yang-Mills theory. Our study builds on previous studies in the literature by considering flavours with different masses. In the supergravity background each family of dynamical flavours produces a spherical cavity with radius proportional to the corresponding quark mass. As a result the geometry has a family of concentric cavities with the density of the flavour branes changing gradually at each cavity, reaching a maximum at the outermost cavity and vanishing inside the innermost cavity. The positive beta function of the flavoured theory is reflected in the grow of the dilaton with the radial distance. The dilaton  blows at some large radial distance corresponding to the Landau pole of the theory. Our solution is uniquely determined by the specification of the position of the Landau pole and the radii of the cavities. 

In the first part of section two we presented our ansatz and the corresponding supergravity equations. In the last part of this section we presented our solution to the supergavity equations and discussed the choice of the integration constants. In the third section of the paper we investigated various properties of our set-up. We began with the derivation of the physical quark masses of each family of dynamical flavours by calculating the energy of a string stretched from the corresponding cavity to the origin of the geometry. In the next subsection we studied the running of the Yang-Mills coupling of the theory, corresponding to the running of the dilaton. We showed how the introduction of an additional family of dynamical flavours increases the beta function of the theory. Finally, in the last part of section three, we investigated the Wilson loop between two cavities, corresponding to two families of light and heavy quarks. We considered different comparison schemes and studied the effect of changing the number of backreacted flavours. As expected we observed a screening of the heavy light potential.

We will close this discussion section by pointing out some directions for future studies. An obvious generalization of this work could be the application of the same philosophy in order to
obtain multicavity solutions for other smeared backreacted backgrounds. The natural next candidate  in this series of backgrounds would be the ABJM (see \cite{Aharony:2008ug} \& \cite{Conde:2011sw, Bea:2013jxa} for the addition of massless and massive flavour D6-branes). The advantage of this construction with respect to the current one is the fact that the 
UV of the theory does not suffer from the presence of a Landau pole. This reflects the fact that D6-branes lift to pure geometry in eleven dimensions, 
so a nice geometrical interpretation of matter is possible. Moving to backgrounds that are closer to realistic theories, we could follow the current philosophy for the field theory that 
is obtained after wrapping $N_c$ D5-branes on a two-cycle of the resolved conifold (see \cite{Maldacena:2000yy} \& \cite{Casero:2006pt, Conde:2011rg} 
for the addition of massless and massive flavour D5-branes). The importance of this construction is that in a geometric setup it successfully encodes confinement and chiral symmetry 
breaking. 

This last construction is also motivated from the recent activity in the front of the holographic computation of the Entanglement entropy \cite{Ryu:2006bv}, and especially 
the generalization of the standard prescription to include non-conformal field theories \cite{Klebanov:2007ws}.
Since the functional forms for both the length of the strip associated to the Entanglement entropy and the length of the Wilson loop are similar, the Entanglement entropy 
and the energy of the Wilson loop as a function of the length, are similar also \cite{Kol:2014nqa}. Including multiple cavities in the Maldacena-Nu\~nez solution 
\cite{Maldacena:2000yy} will imply a confining model that presents multiple first order phase transitions for the Wilson loop. It would be interesting to study the 
effect of this phenomenon on the Entanglement entropy of such a background.


\section*{Acknowledgements}

D.~Z.~is funded by the FCT fellowship SFRH/BPD/62888/2009.
Centro de F\'{i}sica do Porto is partially funded by FCT through the project
CERN/FP/123599/2011. V.~F.~is grateful to the Mainz Institute for Theoretical Physics (MITP) for its hospitality and its partial support during the completion of this work.


\appendix

\section{Analysis of the kappa symmetry for the flavour D7-brane}
\label{app:kappa}

In this part of the appendix we will determine the profile of the supersymmetric embedding for the flavour D7-brane. 
For that we need to specify the kappa symmetry matrix and impose the following condition
\begin{equation} \label{kappa_condition}
\Gamma _{\kappa} \, \epsilon \, = \, \epsilon \,,
\end{equation}
where $\epsilon$ is a killing spinor of the background  \eqref{10dmetric}. 
The form of the kappa symmetry matrix for a Dp-brane in a type IIB theory is
\begin{equation} \label{G-def}
\Gamma_{\kappa} \, = \,{1 \over \sqrt{-\det g }}\, (i\sigma_2)\,\Gamma_{(0)} \, ,
\end{equation}
where $\Gamma_{(0)}$ is denoted as
\begin{equation} \label{G0}
\Gamma_{(0)}={1\over (p+1)!}\,\,\epsilon^{a_1\cdots a_{p+1}}\, \, 
\gamma_{a_1\cdots a_{p+1}}\,,
\end{equation}
with $\gamma_{a_1\cdots a_{p+1}}$ the antisymmetrized product of the induced gamma matrices.

\noindent
We choose the following worldvolume coordinates for the D7-brane\footnote{We have already changed variables from $\sigma$ to $\rho$ as it is explained in 
section \ref{sec:solvingBPS}, namely $d \rho = S^4 d \sigma$ .}
\begin{equation}
\xi^{\alpha}\,=\,(x^{\mu}, \rho, \theta, \varphi, \psi) 
\end{equation}
and constrain our analysis to configurations that $\tau$ and $\chi$ depend on the worldvolume coordinates in the following way 
\begin{equation}
\tau \,=\,{\rm constant} \qquad \& \qquad \chi  \, = \, \chi(\rho) \, . 
\label{chi-ansatz}
\end{equation}
Using  \eqref{kappa_condition} together with the set of projections  \eqref{projections} we will be able to determine the exact form of the function $\chi(\rho)$. 
The induced gamma matrices have the following expressions
\begin{eqnarray}
&& \gamma_{x^{\mu}}\,=\,h^{-{1\over 4}}\Gamma_{\mu}\, ,
\quad \quad
\gamma_{\rho}\,=\,h^{{1\over 4}}\,\Big[\,F \,\Gamma_4\,+\,\frac{1}{2} \, \sin\theta \, \chi'\,\Gamma_5\,\Big]\, ,
\nonumber \\ [12pt]
&& \gamma_{\theta}\,=\,{1 \over 2}\, \cos \frac{\chi}{2}\,S \, h^{{1\over 4}} \,\Gamma_6\, ,
\quad \quad
\gamma_{\psi}\,=\,\, \frac{1}{4} \, h^{{1\over 4}}\, \Big[ 2 \, \cos^2 \frac{\chi}{2} \, F \, \Gamma_9\, + \, \sin \chi \, S\, \Gamma_8 \Big] \, , 
\\ [12pt]
&&\gamma_{\varphi}\,=\,{1 \over 4} \,h^{{1\over 4}}\,
\Big[\,2 \, \cos \theta \, \cos^2 \frac{\chi}{2} \, F \, \Gamma_{9}\,+ \, 2 \, \sin \theta \, \cos \frac{\chi}{2}\,S \, \Gamma_7\, + 
\,\cos \theta \, \sin \chi \, S\, \Gamma_8\,\Big]\,.
\nonumber 
\end{eqnarray}
Following the definition for  $\Gamma_{(0)}$ in \eqref{G0} we have 
\begin{eqnarray}
\Gamma_{(0)} & = & \frac{1}{16} \, \sin \theta \, \cos^3 \frac{\chi}{2} \, S^2 \, \Gamma_{0123} \, \Big[ \,  2 \, \sin \frac{\chi}{2} \, F\, S \, \Gamma_{4867} \, + \, 
\sin \frac{\chi}{2} \, S^2\ \,  \chi'  \, \Gamma_{5867} \, + \, 
\nonumber \\ [12pt]
&&  \qquad  \qquad   \qquad \qquad  \qquad \quad
2 \, \cos \frac{\chi}{2} \, F^2 \, \Gamma_{4967} \, + \, \cos \frac{\chi}{2} \, F \, S \, \chi'  \, \Gamma_{5967} \, \Big] \,.
\end{eqnarray}
Using the projections of \eqref{projections} together with the two relations
\begin{equation}
\Gamma_{4967}\,\epsilon\,=\,-\Gamma_{5867}\,\epsilon\,=\,\epsilon\,\,,
\qquad\qquad
\Gamma_{4867}\,\epsilon\,=\,\Gamma_{5967}\epsilon\,=\, i \, \Gamma_{59}\, \sigma_2 \, \epsilon\, ,
\end{equation}
that easily follow from \eqref{projections}, we obtain the precise way that $\Gamma_{(0)}$  acts on the killing spinor 
\begin{equation} \label{G0-v2}
\Gamma_{(0)}\,\epsilon =  \frac{1}{16} \, \cos^4 \frac{\chi}{2}\, \sin \theta \, S^2 \, 
\Bigg[ F \, S \, \left( \chi' \, + \, 2 \, \tan \frac{\chi}{2} \right) \, \Gamma_{59} +  i \, F^2 \, \left(  \chi' \, \tan \frac{\chi}{2} \, \frac{S^2}{F^2} \, - \, 2 \right) \sigma_2 \,\Bigg]\epsilon\,.
\end{equation}
Since we want to fulfill the condition \eqref{kappa_condition}, the term that is proportional to $ \Gamma_{59}$ should vanish.
In this way we obtain the differential equation for $\chi(\rho)$
\begin{equation} \label{chi-dif}
{d \chi\over d\rho}\,=\,- \, 2 \, \tan \frac{\chi}{2} \,.
\end{equation}
Substituting \eqref{chi-dif} to the determinant of the induced metric we obtain
\begin{equation} \label{det-BPS}
\sqrt{-\det g}_{|\,BPS}\,=\, \frac{1}{8} \, \cos^4 \frac{\chi}{2}\, \sin \theta\, F^2  \, S^2 \, \left( 1 \, + \tan^2 \frac{\chi}{2} \, \frac{S^2}{F^2} \right) \,.
\end{equation}
Therefore, from \eqref{G0-v2}, \eqref{chi-dif}, \eqref{det-BPS} and \eqref{G-def} (for $p$=7), the kappa symmetry condition \eqref{kappa_condition} is satisfied. 
Integrating of \eqref{chi-dif} will give us the profile of $\chi$ as a function of $\rho$ 
\begin{equation}
\chi(\rho)\,=\, 2 \arcsin {e^{ \rho_q} \over e^{\rho}} \, , 
\end{equation}
where $\rho_q$ is the position of the cavity.  Notice that when $\rho_q \rightarrow - \infty$, corresponding to the massless case, the angle $\chi$ goes to zero.


\section{Brane distribution function}
\label{app:BDF}

In this section we will determine the brane distribution function of every cavity $p_{\kappa}(\sigma)$ 
following the analysis of \cite{Conde:2011rg,Conde:2011sw}.  In order to perform this computation we will compare the action of the full set of $N_f$ flavour branes with the 
one that corresponds to the representative embedding.  Since supersymmetry guarantees that both the DBI and WZ parts of the action are the same, we can chose either of them.

The smeared WZ action is coming from the following expression
\begin{equation} \label{smeared-WZ}
S_{WZ}^{smeared} \, = \, T_7 \, \int_{{\cal M}_{10}} \, C_8 \, \wedge \, \Omega_2  \, ,
\end{equation}
where $\Omega_2$ is the smearing form
\begin{equation} \label{smearing}
\Omega_2 \, = \,  - \, d F_1 \, = \, - \,\sum_{\kappa}  Q_f^{\kappa} \, p_{\kappa}' \, d \sigma \wedge (d\tau \, + \, A_{CP^2} )  \, 
- \,\sum_{\kappa} Q_f^{\kappa} \, p_{\kappa} \, d A_{CP^2}  \, , 
\end{equation}
and $C_8$ is the potential that is coming from the Hodge dual of $F_1$ in the standard way, 
\begin{equation}
F_9 = e^{2 \Phi} \star F_1 \, .
\end{equation} 
The explicit expression for the potential may come from the following calibration condition
\begin{equation} \label{C8_potential}
\star F_1 \, = \, e^{-2 \Phi} \, d (e^{\Phi} \, {\cal K}) \quad \Rightarrow \quad 
C_8 \, = \, e^{\Phi} \, {\cal K} \, ,
\end{equation} 
where ${\cal K}$ is the calibration form 
\begin{equation} \label{calibration}
{\cal K} \, = \,  - \, \frac{1}{2} \, \,h^{- 1} \, \, d^4 x  \, \wedge \,  J_2 \, \wedge \, J_2 \, ,
\end{equation}
and $J_2$ the following two-form
\begin{equation} 
J_2 \, = \, e^{4} \, \wedge \,  e^{9} \, - \,  e^{5} \, \wedge \,  e^{8} \, - \,  e^{6} \, \wedge \,  e^{7}\, .
\end{equation}
Combining \eqref{smeared-WZ}, \eqref{smearing} \& \eqref{C8_potential} and integrating over
the angles of the sphere ($\chi, \theta, \varphi, \psi$ \& $\tau$) we obtain the following expression
for the smeared WZ action
\begin{equation} \label{lag_smeared}
{\cal L}_{WZ}^{smeared} \, = \, \frac{1}{2} \,\pi^2\,T_{7}\, e^{\Phi} \, S^4 \,\sum_{\kappa}N_f^{\kappa}\,
\Big[p_{\kappa}' \, + \, 4 \, p_{\kappa} \, F^2 \, S^2 \Big] \,.
\end{equation}

Multiplying the lagrangian for a single massive embedding\footnote{that wraps $\theta, \varphi, \psi$, 
sits at fixed $\tau$ and has a profile along $\chi = \chi(\sigma)$} with $N_f^{\kappa}$ we get
\begin{equation} \label{lag_massive}
{\cal L}_{WZ} \, = \, \frac{1}{4} \,\pi^2 \,T_{7}\, e^{\Phi} \, S^4\,\sum_{\kappa}N_f^{\kappa}\,
\Bigg[8 \, F^2 \, S^2 \, \cos^4 \frac{\chi_{\kappa}}{2}\, + \, 2 \, 
\partial_{\sigma} \left( \cos^4 \frac{\chi_{\kappa}}{2} \right) \Bigg] \,.
\end{equation}
Equating the coefficients in front of the charges $N_f^{\kappa}$ in both equations \eqref{lag_smeared} \& \eqref{lag_massive}, we obtain the differential equation for the
brane distribution function for every cavity
\begin{equation}
\partial_{\sigma} \Bigg[ p_{\kappa} \, - \, \cos^4 \frac{\chi_{\kappa}}{2} \Bigg] \, + \, 
4 \, F^2 \, S^2 \, \Bigg[ p_{\kappa} \, - \, \cos^4 \frac{\chi_{\kappa}}{2} \Bigg] \, , 
\end{equation}
which is trivially solved 
\begin{equation} \label{p_function}
p_{\kappa}(\sigma) \, = \, \cos^4 \frac{\chi_{\kappa}(\sigma)}{2} \, .
\end{equation}


\end{document}